\documentclass[]{spie}  

\usepackage{amsmath,amsfonts,amssymb}
\usepackage{graphicx}
\usepackage[colorlinks=true, allcolors=blue]{hyperref}

\title{Testing the limits of human vision with quantum states of light: past, present, and future experiments}

\author[a]{Rebecca M. Holmes}
\author[b]{Michelle M. Victora}
\author[c]{Ranxiao Frances Wang}
\author[b]{Paul G. Kwiat}
\affil[a]{Los Alamos National Laboratory, P.O. Box 1663, Los Alamos, NM 87545 USA}
\affil[b]{Department of Physics, University of Illinois at Urbana-Champaign, 1110 West Green St, Urbana, IL USA 61801-3003}
\affil[c]{Department of Psychology and Beckman Institute, University of Illinois at Urbana-Champaign, 603 E. Daniel St, Champaign, IL USA 61820}

\begin{document} 
\maketitle

\begin{abstract}
The human eye contains millions of rod photoreceptor cells, and each one is a single-photon detector. Whether people can actually see a single photon---which requires the rod signal to propagate through the rest of the noisy visual system and be perceived in the brain---has been the subject of research for nearly 100 years. Early experiments hinted that people could see just a few photons, but classical light sources are poor tools for answering these questions. Single-photon sources have opened up a new area of vision research, providing the best evidence yet that humans can indeed see single photons, and could even be used to test quantum effects through the visual system. We discuss our program to study the lower limits of human vision with a heralded single-photon source based on spontaneous parametric downconversion, and present two proposed experiments to explore quantum effects through the visual system: testing the perception of superposition states, and using a human observer as a detector in a Bell test.
\end{abstract}


\section{Introduction}
\label{intro}

The quantum optics community has long been interested in the human visual system, which is likely to be sensitive to single photons. Early experiments were limited by the randomness of classical light sources, but the era of true single-photon sources and tunable photon statistics has enabled new areas of research, including measuring the quantum efficiency of a rod photoreceptor cell (about 33\%)\cite{Phan2014} and measuring photon statistics of different light sources using a rod cell as a sensor \cite{Sim2012}. A recent experiment provided the best proof to date that the visual system can detect a single photon \cite{Tinsley2016}, and others have explored temporal integration in the visual system at the few-photon level \cite{Holmes2017}.

These advances in single-photon vision research provide a unique opportunity to study quantum effects through the visual system, including superposition and entanglement. This paper will give a short overview of previous research on single-photon vision and current capabilities, and describe two proposed experiments to study the perception of superposition states and to use a human observer as a detector in a Bell test. 

\section{Classic experiments}
\label{sec:classic} 

Shortly after physicists began to think of light as photons in the early twentieth century, it became clear that the statistics of individual photons were likely to be important to understanding the extreme lower limit of human vision \cite{Bouman2012}. One of the earliest and best-known experiments on the lower limit of vision was done by Hecht, Schlaer, and Pirenne in 1942 \cite{Hecht1942}.

\begin{figure} [ht]
   \begin{center}
   \begin{tabular}{c} 
   \includegraphics[width=0.6 \textwidth]{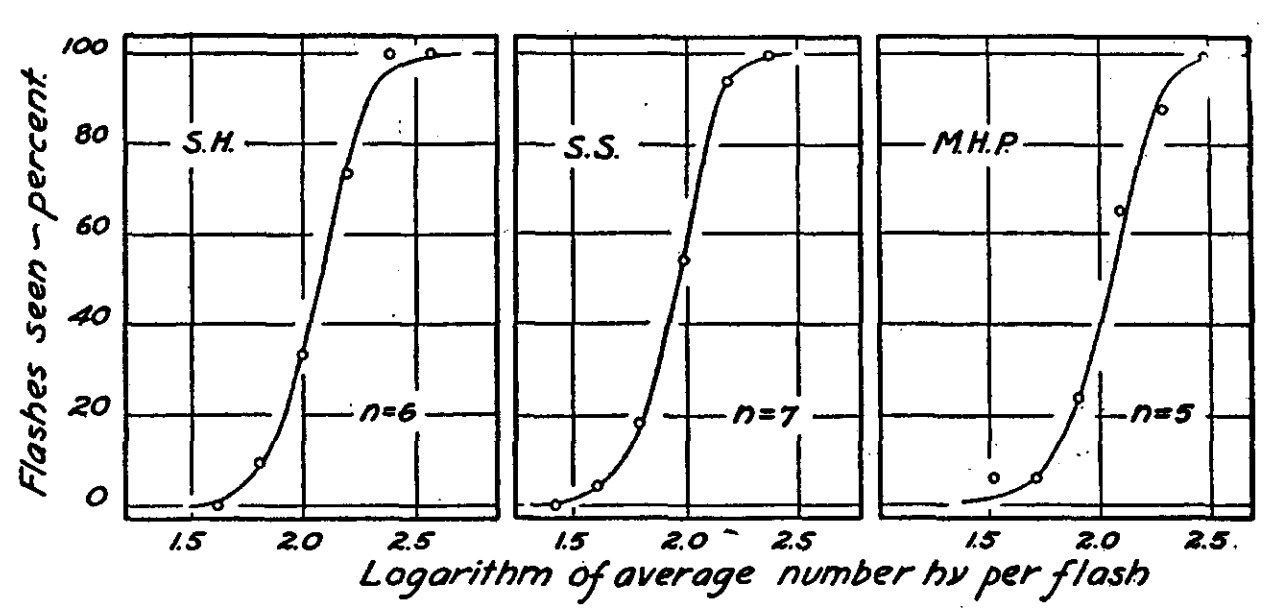}
   \end{tabular}
   \end{center}
   \caption[example] 
   { \label{fig:hecht} 
Data from Hecht et al.\cite{Hecht1942} Fitting a Poisson model to the relationship between the mean number of photons in a flash and the frequency with which an observer reports that it was visible gives an estimate of the visual threshold $n$.}
   \end{figure}

Hecht et al. presented human observers with very dim flashes of light, with mean photon numbers between 50 and 400. After each flash, the observer (each of the study's three co-authors) was asked whether it was visible or not. The mean photon number of the flashes was varied, and it was determined how often the observers detected the flash at each level. With the assumption that the number of photons detected by the visual system in each trial was a Poisson-distributed random variable, and that there was some threshold number of photons $n$ required for perception, Hecht et al. estimated that the threshold for vision was between 5 and 7 photons, depending on the observer (Figure \ref{fig:hecht}).

This experiment provided some of the first clear evidence that the rod cells can detect single photons: the flashes of light were incident on an area containing about 500 rods, so if just 5-7 photons were visible, any individual rod cell would be unlikely to detect more than one. However, there are several problems with this experimental design which made Hecht et al.'s conclusion that vision requires 5-7 photons likely to be an overestimate. Most importantly, asking observers to simply report whether a flash was seen or not can introduce an artificial threshold above the true lower limit of vision, because observers may have a bias against making false positive responses.

\begin{figure} [ht]
   \begin{center}
   \begin{tabular}{c} 
   \includegraphics[width=0.8\textwidth]{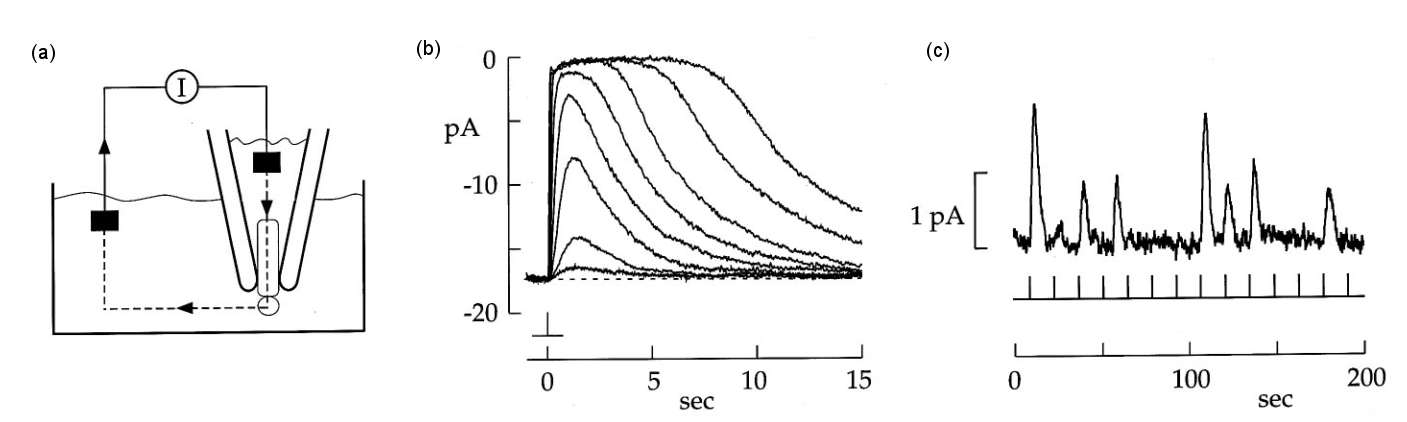}
   \end{tabular}
   \end{center}
   \caption[example] 
   { \label{fig:electrode} 
(a) Measuring the photocurrent of a single rod cell. The rod is held in a pipette by suction, so that the light-induced current flows through an electrode placed in the surrounding electrolyte bath. (b) Photocurrents from a monkey rod cell in response to brief flashes at t = 0. The smallest peak corresponds to a flash with a mean photon absorption number of $\sim$1, the largest to $\sim$500. (c) Photocurrent pulses produced by a series of dim flashes. The rod produces currents of different strengths for absorptions of 1 or 2 photons. Figures modified from \citenum{Rieke1998}, data from \citenum{Baylor1984}.}
   \end{figure}

Indeed, later experiments which instructed observers to rate the visibility of a dim flash of light on a scale from 0-6 found a lower threshold for vision, perhaps as low as one photon (for some observers) \cite{Sakitt1972}. In vitro measurements of individual rod cells also showed that the cells produce discrete electrical signals in response to dim flashes of light, with the smallest signals seeming to correspond to single photons \cite{Rieke1998} (see Figure \ref{fig:electrode}).

However, all these experiments were limited by the randomness of classical light sources, which can never produce guaranteed single photons. The development of single-photon sources created new possibilities in vision research, as discussed in Section \ref{single-photon}.

\section{The single-photon era}
\label{single-photon}

Single photon sources were developed for quantum optics and quantum information research, and include single atoms \cite{McKeever2004}, nitrogen-vacancy centers in diamond \cite{Kurtsiefer2000,Beveratos2001}, quantum dots \cite{Michler2000}, and spontaneous parametric downconversion (SPDC) photon pair sources \cite{Hong1986}. SPDC sources are in many ways ideal for single-photon vision research, as they can be very bright, can produce light at a broad range of wavelengths (the rod cells are most sensitive around 500 nm), and can achieve high efficiency limited primarily by optical loss. With some modification they can also readily produce photon pairs entangled in polarization and other degrees of freedom \cite{Kwiat1999}.

\begin{figure} [ht]
   \begin{center}
   \begin{tabular}{c} 
   \includegraphics[width=0.7\textwidth]{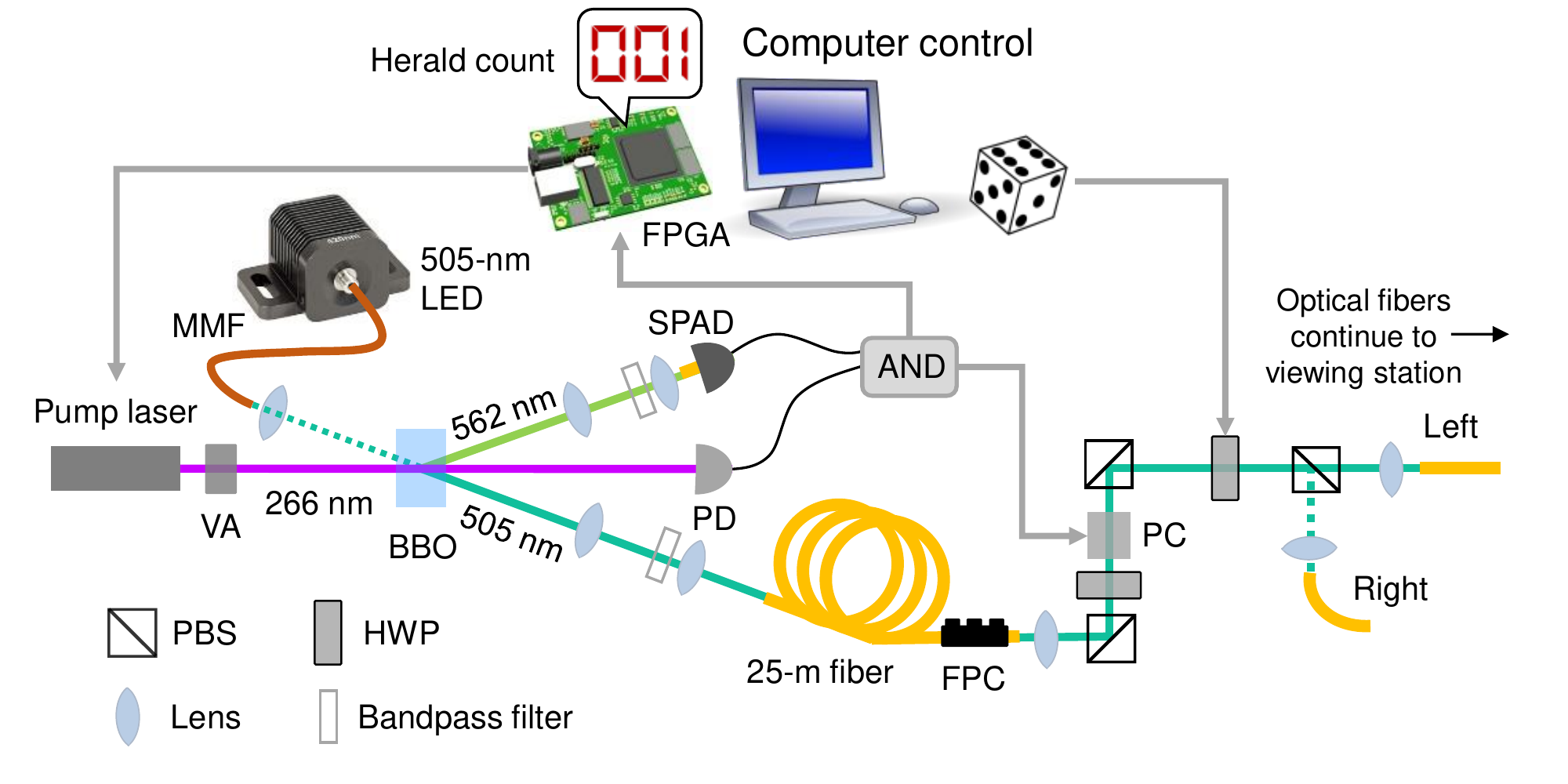}
   \end{tabular}
   \end{center}
   \caption[example] 
   { \label{fig:sps} 
Schematic of our single-photon source. The pump laser power is controlled by a variable attenuator (VA). It produces pairs of photons at 562 and 505~nm inside a nonlinear crystal (BBO); the 562-nm herald photons are sent to a single-photon avalanche photodiode (SPAD) and counted (in coincidence with a photodiode triggered on the pump laser to eliminate background counts) by an FPGA. The 505-nm photons travel through a 25-m fiber for optical delay, and their polarization is matched to a polarizing beam splitter (PBS) with a fiber polarization controller (FPC); herald detections then trigger a Pockels cell (PC) which allows heralded 505-nm photons to be reflected by a second PBS. When a predetermined herald count is reached, the pump laser is shut off. The 505-nm signal photons also pass through a computer-controlled half-wave plate (HWP) and a third PBS, which allows them to be directed into one of two output fibers connected to an observer viewing station (Figure \ref{fig:viewing-station}). A 505-nm LED can be coupled into the path of the signal photons as an alignment aid for an observer.}
   \end{figure}

An example SPDC pair source, developed in our lab and optimized for human vision research \cite{Holmes2012,Holmes2015}, is shown in Figure~\ref{fig:sps}. The heralding efficiency of this source (the probability that a photon is sent to an observer when a herald photon is detected) was measured to be 38.5\%. At the lowest energy per pulse (80-kHz rep rate, herald detection rate of $\sim$52~Hz) the $g^{(2)}(0)$ was measured to be $0.0023$. It produces single photons at 505 nm, near the peak of the rods' spectral sensitivity. 

\begin{figure} [ht]
   \begin{center}
   \begin{tabular}{c} 
   \includegraphics[width=0.7\textwidth]{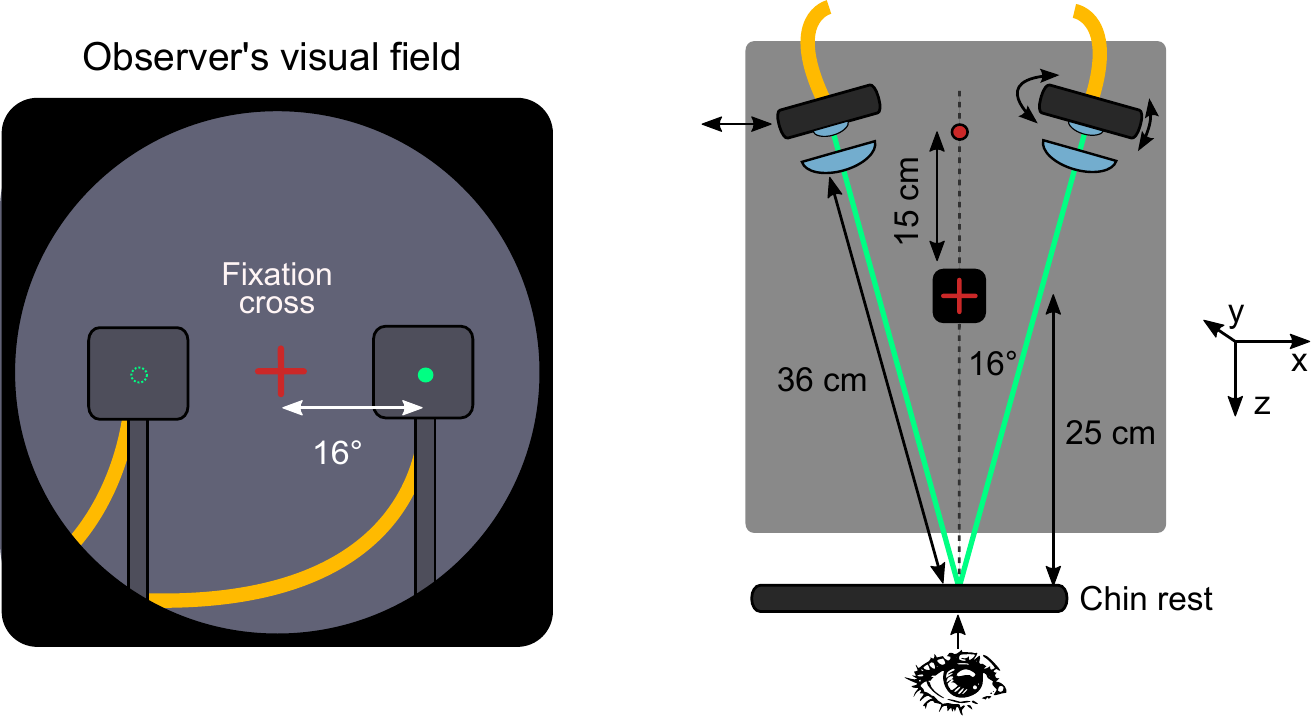}
   \end{tabular}
   \end{center}
   \caption[example] 
   { \label{fig:viewing-station} 
The observer's visual field and a top-down schematic of our observer viewing station. The fixation cross (not shown to scale in visual field) consists of a dim 700-nm LED behind a crosshairs mask (the rod cells are not sensitive to far-red wavelengths). The left and right beams are both aligned to the observer's right eye when the observer is positioned in a chin rest.}
   \end{figure}

Human vision research also requires a viewing station to deliver photons to the observer (Figure \ref{fig:viewing-station}). Ours is designed to deliver photons to one or both of two spatially separated spots on the retina, corresponding to positions at about $\pm 16$ degrees to the left and right of the fovea. This feature allows us to use an improved experimental design compared to Hecht et al.: instead of asking whether a photon was seen or not, we can randomly (using the half-wave plate and polarizing beam splitter shown in Figure \ref{fig:sps}) send a photon to either the left or the right spot and ask \emph{where} it was seen. This removes the artificial threshold effect that may occur when the observer is asked to judge whether a stimulus was present or not. If the observer is able to choose ``left'' or ``right'' with accuracy significantly above 50\%, we can conclude that they were able to see the stimulus. The most significant drawback of this approach is that due to the relatively high optical loss in the eye (estimated to be 90-97\%), very large numbers of time-consuming experimental trials are typically required to demonstrate an effect, as in most trials the observer does not actually detect a photon.

We have used this source to study how the visual system integrates photon detections that occur within a short time window\cite{Holmes2017}, and others used a similar SPDC source and experimental design (with left and right replaced by earlier and later photon delivery times) to show that observers could achieve accuracy above 50\% for single photons. A follow-on study with a much larger number of experimental trials and improved experimental design (incorporating an equal number of control trials in which no photon is present) will be valuable to confirm this result. However, we consider it likely that humans can indeed detect single photons. An exciting possibility is that a similar single-photon source could be now used to explore quantum effects through the visual system. Two proposed experiments are presented in Section \ref{quantum}. 
   
\section{Proposed experiments: superposition and entanglement}
\label{quantum}

If humans can detect single photons, then a wide range of exciting future work and experiments opens up to us. By examining how a human observer interacts with and measures quantum phenomena, we can test the predictions of standard quantum mechanics and even give a human observer a direct role in a test of local realism.

\subsection{Perception of superposition states}

   \begin{figure} [ht]
   \begin{center}
   \begin{tabular}{c} 
   \includegraphics[width=0.7\textwidth]{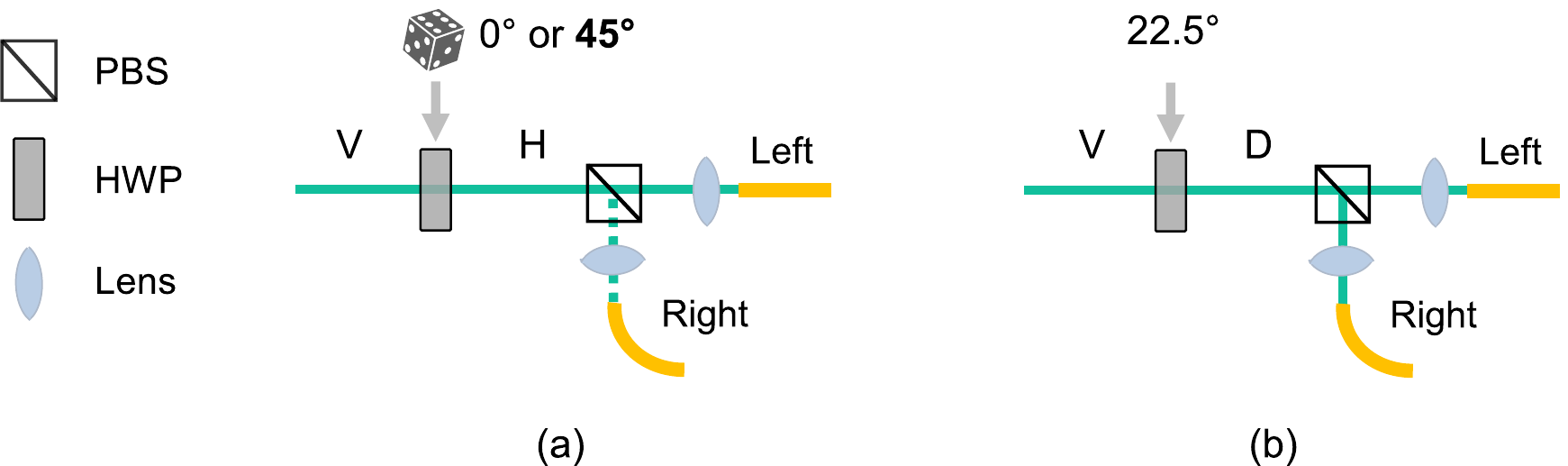}
   \end{tabular}
   \end{center}
   \caption[example] 
   { \label{fig:superposition} 
(a) Left/right switch as it is used in a single-photon vision test. The half-wave plate (HWP) is randomly set to either 0$^{\circ}$ or 45$^{\circ}$, which directs the photon to either the left or the right fiber. The photon is in a classical mixture of the left and right positions. (b) Modification to create a superposition state. The HWP is set to 22.5$^{\circ}$, which rotates vertical (V) polarization to diagonal (D) polarization, an equal, coherent superposition of H and V. The polarizing beam splitter (PBS) couples polarization to the left and right path, creating a superposition of left and right. Note that in this case, we (the researchers) do not know which side the photon will be detected on.}
   \end{figure}

One relatively easy test we can perform is to determine whether humans perceive any difference between a photon in a superposition state and a classical mixed state. A superposition experiment in the visual system has been of great interest for many years, and several approaches have been proposed \cite{Ghirardi1999,Thaheld2003}. To carry out a version of this experiment, we can use our setup from Figure \ref{fig:sps}, but in addition to experimental trials which present a single photon on either the left or the right side of the retina, we also randomly present trials with a photon in a superposition of the left and right spots. This is easily achieved by rotating the half-wave plate shown in Figure \ref{fig:superposition} to 22.5$^{\circ}$ to produce the state
\begin{equation}\label{superposition}
\frac{1}{\sqrt{2}}(|H, \textrm{right}\rangle + |V, \textrm{left}\rangle)
\end{equation}
As in a single-photon vision test, the observer is asked to indicate on which side the photon was seen in each trial. According to standard quantum mechanics, there should be no difference in perception between an equal superposition and an equal classical mixture. Any statistically significant difference in the proportion of left and right responses between the two conditions (after carefully accounting for any bias in the experimental apparatus) would be evidence for an unexpected effect, and could have implications for alternative interpretations of quantum mechanics (e.g., macrorealism\cite{Ghirardi1986a,Leggett2002}).
   
\subsection{Bell test with a human observer}

   \begin{figure} [ht]
   \begin{center}
   \begin{tabular}{c} 
   \includegraphics[width=0.7\textwidth]{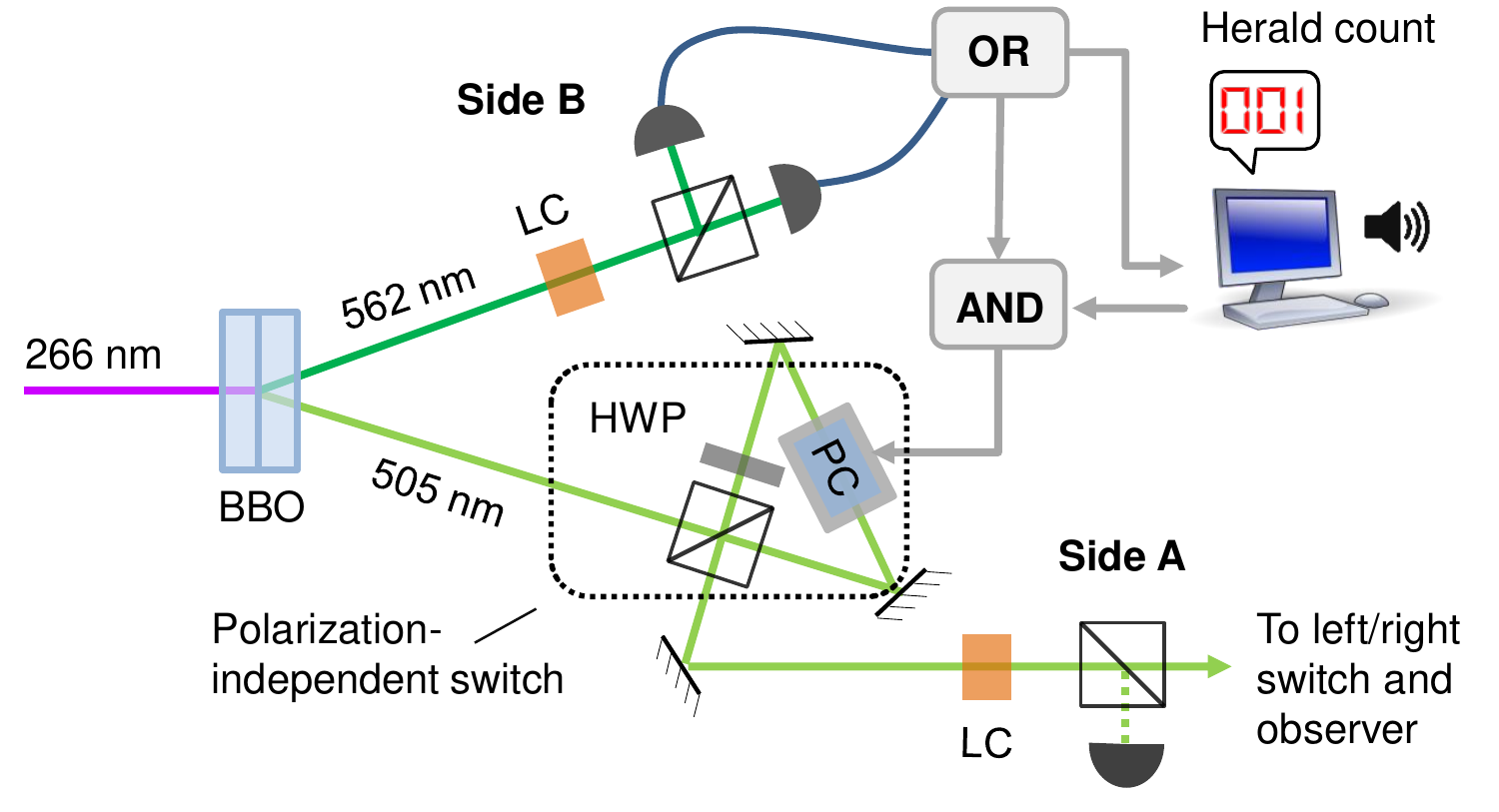}
   \end{tabular}
   \end{center}
   \caption[example] 
   { \label{fig:entanglement} 
Simplified schematic of a Bell test with one detector replaced by a human observer. When one of the detectors on side A indicates a photon has been measured with setting $b' = 67.5^{\circ}$, an audio signal alerts the observer to watch for a photon on their side. A polarization-independent switch prevents unheralded photons: if the Pockels cell (PC) is not activated, the half-wave plate (HWP) rotates the polarization of a incoming photon, and it returns to the source crystal. If the PC is activated by a herald detection, it cancels the effect of the HWP and the photon is sent to the observer (with its initial polarization, and entanglement, unchanged). Liquid crystals (LC) are used to set the measurement bases on each side.}
   \end{figure}

Another exciting experiment we could perform is a test of local realism using a human observer. As a first step, we would only replace one detector with a human observer, and the other measurements would use our high-efficiency photon counters (see Figure \ref{fig:entanglement}). By simultaneously pumping two orthogonal nonlinear crystals, we can produce polarization-entangled pairs of photons \cite{Christensen2013, Kwiat1999, Altepeter2005a}. We then use the well-known CH inequality \cite{Clauser1974a} which relates coincidence probabilities ($c$) and singles probabilities ($s$) for settings $a, a', b, b'$ on polarization analyzers $A$ and $B$: 
\begin{equation}\label{ch}
c_{12}(a,b) + c_{12}(a',b) + c_{12}(a',b') - c_{12}(a,b') \leq s_{1}(a') + s_{2}(b)
\end{equation}
It can be shown that any local realistic theory must obey this inequality. In an initial experiment, we use single-photon detectors to measure all terms except $c_{12}(a',b')$. Using optimal analysis choices ($a = 0^{\circ}$, $a' = 45^{\circ}$, $b = 22.5^{\circ}$, and $b' = 67.5^{\circ}$), the inequality becomes
\begin{equation}
3 \cos^{2}(22.5^{\circ})/2 - p_{\text{obs}} \leq 1
\end{equation}
\begin{equation}
 p_{\text{obs}} \geq 0.28
\end{equation}
where $p_{\text{obs}}$ is the probability that the observer detects a photon on their side ($A$) with measurement setting $a'$ ($45^{\circ}$) when a photon is detected on side $B$ with measurement setting $b'$ ($67.5^{\circ}$). Thus, if $p_{\text{obs}}$ exceeds 0.28 with statistical significance, the CH inequality is violated.

A forced-choice design, similar to a single-photon vision test, can be used to control for the low probability that an observer will detect a photon in any given trial. If the measurement on side $B$ indicates the desired outcome for the $c_{12}(a',b')$ term, the entangled photon continues to a left/right switch (like the one shown in Figure~\ref{fig:superposition}a) and is randomly directed to one side of the observer's visual field. Additionally, a non-entangled photon is delivered to the other side of the visual field with 28\% probability. The observer makes independent judgments about whether a photon was present on each side. If $p_{\text{obs}} = 0.28$, the observer will see the side with the entangled photon as often than they see the control (non-entangled) side. If they see the entangled side significantly more often than the control, the measurement outcome violates the CH inequality. Note that the ``timing'' and ``detection'' loopholes are unlikely to ever be accounted for in such a test. However, it would still be a unique and interesting experimental challenge.

\section{Conclusions}

Now that experiments with single-photon sources have shown that humans are likely to be able to detect single photons, a wide range of interesting new experiments in both physics and psychology may be proposed. This paper has given an overview of previous work on the lower limit of human vision, and presented two possible experiments to test quantum mechanics through the visual system, including superposition states and entanglement. The primary challenge for these and other single-photon vision experiments will be the low probability that a photon is transmitted to the photoreceptors and detected in any given experimental trial (perhaps 5-10\%, assuming a perfectly efficient source), and the corresponding requirement for a very large number of experimental trials.

While the human observer makes these proposed experiments unique and interesting, we emphasize that we do not propose to test whether the presence of a conscious observer plays a role in the outcome of these experiments; rather, these experiments take advantage of the unique framework of the visual system to test the predictions of quantum mechanics, and may even be able to provide experimental limits on alternative proposals such as macrorealism. 

There are also a number of interesting studies in psychophysics that could make use of our single-photon source. We could investigate other aspects of temporal summation at the lowest light levels, such as whether a dim light is ever perceived to be quantized. Using deformable mirrors and spatial light modulators, we could study spatial summation by varying the size of a few-photon stimulus on the retina. More advanced sources that can produce higher photon number states \cite{McCusker2009a} could also be used to measure the visual sensitivity function for exact photon numbers.

\bibliography{library} 
\bibliographystyle{spiebib} 

\end{document}